\documentclass[10pt,cite,aps,pra,amsmath,amsfonts,amssymb,superscriptaddress,showpacs,twocolumn]{revtex4}
\renewcommand{\vec}[1]{\mathbf{#1}}
\usepackage{graphics}
\usepackage{graphicx}
\usepackage{wasysym}
\usepackage{setspace}
\usepackage{color}
\usepackage{nicefrac}
\usepackage{braket}

\newcommand{\figref}[1]{Fig.~\ref{fig:#1}}
\newcommand{\Figref}[1]{Figure~\ref{fig:#1}}

\renewcommand{\eqref}[1]{Eq.~(\ref{eq:#1})}

\newcommand{\citeasnoun}[1]{Ref.~\onlinecite{#1}}
\newcommand{\secref}[1]{Sec.~\ref{sec:#1}}

\newcommand{\bb}[1]{\mathbb{#1}}

\def\a{s}
\def\b{s}
\newcommand{\add}[1]{\if\a\b{{\color{red} #1}}\else{#1}\fi}
\newcommand{\comm}[1]{\if\a\b{{\color{blue}\{\small \sc #1\}}}\else{}\fi}
\newcommand{\del}[1]{{\if\a\b{{\color{magenta}[[#1]]}}\else{}\fi}}

\begin{document}

\title{Casimir micro-sphere diclusters and three-body effects in
  fluids}

\author{Jaime Varela}
\affiliation{Department of Physics,
Massachusetts Institute of Technology, Cambridge, MA 02139}
\author{Alejandro W. Rodriguez}
\affiliation{School of Engineering and Applied Sciences,
Harvard University, Cambridge, MA 02139}
\author{Alexander P. McCauley}
\affiliation{Department of Physics,
Massachusetts Institute of Technology, Cambridge, MA 02139}
\author{Steven G. Johnson}
\affiliation{Department of Mathematics,
Massachusetts Institute of Technology, Cambridge, MA 02139}

\begin{abstract}
  Our previous article [Phys. Rev. Lett. \textbf{104}, 060401 (2010)]
  predicted that Casimir forces induced by the material-dispersion
  properties of certain dielectrics can give rise to stable
  configurations of objects. This phenomenon was illustrated via a
  dicluster configuration of non-touching objects consisting of two
  spheres immersed in a fluid and suspended against gravity above a
  plate. Here, we examine these predictions from the perspective of a
  practical experiment and consider the influence of non-additive,
  three-body, and nonzero-temperature effects on the stability of the
  two spheres.  We conclude that the presence of Brownian motion
  reduces the set of experimentally realizable silicon/teflon
  spherical diclusters to those consisting of layered micro-spheres,
  such as the hollow-core (spherical shells) considered here.
\end{abstract}

\maketitle

\section{Introduction}

\begin{figure}[t]
\includegraphics[width=0.7\columnwidth]{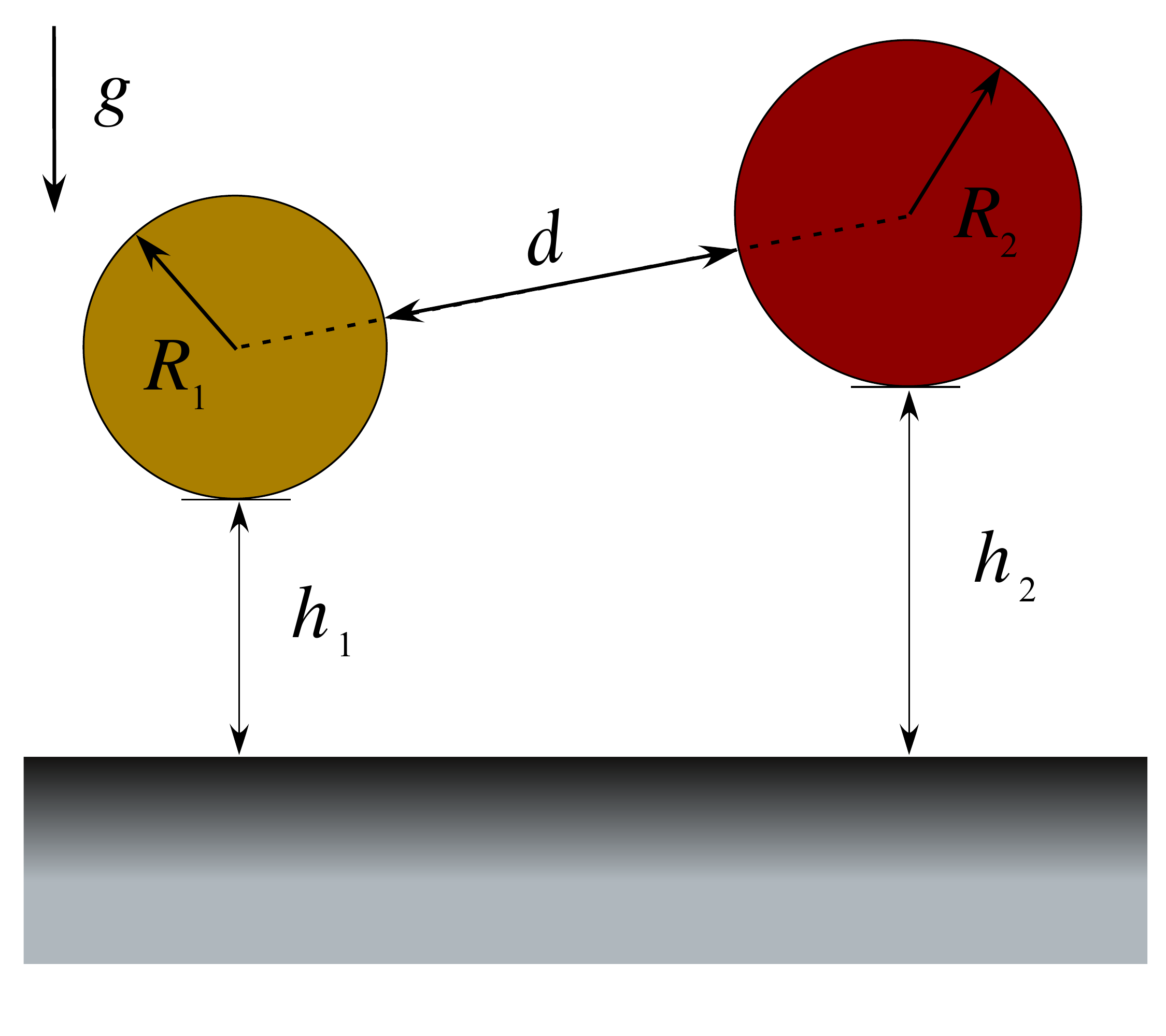}
\caption{Schematic of two-sphere dicluster geometry consisting of two
  dielectric spheres of radii $R_1$ and $R_2$ separated by a
  center--center distance $d$ from each other, and suspended by
  heights $h_1$ and $h_2$, respectively, above a dielectric plate.}
\label{fig:fig1}
\end{figure}

In this paper, we investigate the influence of non-additive/three-body
and nonzero-temperature effects on our earlier prediction that the
Casimir force (which arises from quantum electrodynamic
fluctuations~\cite{casimir,Lifshitz80,milton04}) can enable dielectric
objects (micro-spheres) with certain material dispersions to form
stable non-touching configurations (diclusters) in
fluids~\cite{RodriguezMc10:PRL,Rodriguez08:PRL}.  Such micro-sphere
interactions are predicted to possess a variety of unusual Casimir
effects, including repulsive
forces,~\cite{Kenneth,Munday09,Dzyaloshinskii61} a strong interplay
with material dispersion~\cite{RodriguezMc10:PRL}, and strong
temperature dependences~\cite{RodriguezW10:PRL}, and may have
applications in microfluidic particle
suspensions~\cite{RahiZa09:arxiv,McCauleyRo10:PRA}.  A typical
situation considered in this paper is depicted in \figref{fig1},
consisting of silicon and teflon micro-spheres suspended in ethanol
above a gold substrate.  Although our earlier work considered pairs of
micro-spheres suspended above a substrate in the additive/pairwise
approximation, summing the exact two--body sphere--sphere and
sphere--substrate interactions, in this paper we perform exact
three-body calculations.  In \secref{pairwise}, we explicitly
demonstrate the breakdown of the pairwise approximation for
sufficiently small spheres, in which an adjacent substrate modifies
the equilibrium sphere separation, but we also identify experimentally
relevant regimes in which pairwise approximations (and even a
parallel-plate proximity-force/PFA approximation~\cite{Derjaguin}) are
valid.  In \secref{temp}, we also consider temperature corrections to
the Casimir interactions.  Although a careful choice of materials can
lead to a large temperature dependence stemming from the thermal
change in the photon
distribution~\cite{Bostrom,Bordag,RodriguezW10:PRL}, we find that such
thermal-photon effects are negligible ($< 2$\%) for the materials
considered here.  However, we show that substantial modifications to
the objects separations occur due to Brownian motion of the
micro-spheres.  This effect can be reduced by lowering the
temperature, limited by the freezing point of ethanol ($T \approx
159$~K ), or by increasing the sphere diameters.  We propose
experimentally accessible geometries consisting of hollow
micro-spheres (which can be fabricated by standard
methods~\cite{Wilcox}) whose dimensions are chosen to exhibit a clear
stable non-touching equilibrium in the presence of Brownian
fluctuations.  We believe that this work is a stepping stone to direct
experimental observation of these effects.

In fluid-separated geometries the Casimir force can be repulsive,
leading to experimental wetting effects~\cite{wetting, Capillarity, Israelachvili}
and even recent direct measurements of the repulsive force in fluids
for sphere-plate geometries~\cite{moh1,Munday07, Feiler}.  In
particular, for two dielectric/metallic materials with permittivity
$\varepsilon_1$ and $\varepsilon_3$ separated by a fluid with
permittivity $\varepsilon_2$, the Casimir force is repulsive when
$\varepsilon_1 < \varepsilon_2 <
\varepsilon_3$~\cite{Dzyaloshinskii61}.  More precisely, the
permittivities depend on frequency $\omega$, and the sign of the force
is determined by the ordering of the $\varepsilon_k(i\kappa)$ values
at imaginary frequencies $\omega=i\kappa$ (where $\varepsilon_k$ is
purely real and positive for any causal passive
material~\cite{Dzyaloshinskii61}).  If the ordering changes for
different values of $\kappa$, then there are competing repulsive and
attractive contributions to the force.  At larger or smaller
separations, smaller or larger values of $\kappa$, respectively,
dominate the contributions to the total force, and so the force can
change sign with separation.  For example, if $\varepsilon_1 <
\varepsilon_2 < \varepsilon_3$ for large $\kappa$ and $\varepsilon_1 <
\varepsilon_3 < \varepsilon_2$ for small $\kappa$, then the force may
be repulsive for small separations and attractive for large
separations, leading to a stable equilibrium at an intermediate
nonzero separation.  Alternatively, for a sphere--plate geometry in
which the sphere is pulled downwards by gravity, a purely repulsive
Casimir force (which dominates at small separations) will also lead to
a stable suspension.  These basic ideas were exploited in our previous
work~\cite{RodriguezMc10:PRL} to design sphere--sphere and
sphere--plate geometries exhibiting a stable non-touching
configuration.  The effects of material dispersion are further
modified by an interplay with geometric effects (which set additional
length-scales beyond that of the separation), as well as by
nonzero-temperature effects which set a Matsubara length-scale $2\pi
kT/\hbar$~\cite{Bordag} that can further interact with dispersion in
order to yield strong temperature corrections~\cite{RodriguezW10:PRL}.
Experimentally, stable suspensions are potentially appealing in that
one would be measuring static displacements rather than force between
micro-scale objects.  The stable configurations may be further
modified, however, by three-body effects in sphere--sphere--plate
geometries and by Brownian motion of the particles within the
potential well created by the Casimir interaction, and these effects
are studied in detail by the present paper.

Until the last few years, theoretical predictions of Casimir forces
were limited to a small set of simple geometries (mainly planar
geometries) amenable to analytical solution, but a number of
computational schemes have recently been demonstrated that are capable
of handling complicated (and, in principle, arbitrary) geometries and
materials~\cite{Rahi09:PRD,Rodriguez07:PRA,Homer09}.  Here, since the
geometries considered in this paper consist entirely of spheres and
planes, we are able to adapt an existing technique~\cite{Rahi09:PRD}
based on Fourier-like (``spectral'') expansions that semi-analytically
exploits the symmetries of this problem.  This technique, formulated
in terms of the scattering matrices of the objects in a basis of
spherical or plane waves, was developed in various forms by multiple
authors~\cite{Emig07,Rahi09:PRD, Kenneth08}, and we employ the
generalization of \cite{Rahi09:PRD}.  Although this process is
described in detail elsewhere~\cite{Rahi09:PRD} and is reviewed for
the specific geometries of this paper in the appendix, the basic idea
of the calculation is as follows.  The Casimir energy can be expressed
via path integrals as an integral $\int_0^\infty \log \det A(\kappa)
d\kappa$ over imaginary frequencies $\kappa$, where $A$ is a
``T-matrix'' related to the scattering matrix of the system.  In
particular, one needs to compute the scattering matrices relating
outgoing spherical waves from each sphere (or planewaves from each
plate) being reflected into outgoing spherical waves (or planewaves)
from every other sphere (or plate), which can be expressed
semi-analytically (as infinite series) by ``translation matrices''
that re-express a spherical wave (or planewave) with one origin in
terms of spherical waves (or planewaves) around the origin of the new
object~\cite{Rahi09:PRD}.  This formalism is exact (no uncontrolled
approximations) in the limit in which an infinite number of
spherical/plane waves is considered.  To obtain a finite matrix $A$,
the number of spherical waves (or spherical harmonics $Y_{\ell m}$) is
truncated to a finite order $\ell$.  Because this expansion converges
exponentially fast for spheres~\cite{Canaguier09, Rahi09:PRD}, we find
that $\ell \leq 12$ suffices for $< 1$\% errors with the geometries in
this paper.  (Conversion from planewaves to spherical waves is
performed by a semi-analytical formula~\cite{Rahi09:PRD} that involves
integrals over all wavevectors, which was performed by a standard
quadrature technique for semi-infinite integrals~\cite{Genz83}.)
Although it is possible to differentiate $\log \det A$ analytically to
obtain a trace expression for the force~\cite{Homer09}, in this paper
we use the simple expedient of computing the energy and
differentiating numerically via spline interpolation.  Previously,
\citeasnoun{Lopez09} employed the same formalism in order to study a
related geometry consisting of vacuum-separated perfect-metal spheres
adjacent to a perfect-metal plate, where it was possible to employ the
method of images to reduce the computational complexity
dramatically. That work found a three-body phenomenon in which the
presence of a metallic plate resulted on a stronger attractive
interaction between the spheres, and that this effect becomes more
prominent at larger separations~\cite{Lopez09}, related to an earlier
three-body effect predicted for cylindrical shapes~\cite{RahiRo07,Rodriguez07:PRL}.
Here, we examine dielectric spheres and plate immersed in a fluid and
therefore cannot exploit the method of images for simplifying the
calculation, which makes the calculation much more expensive because
of the many oscillatory integrals that must be performed in order to
convert between planewaves (scattering off of the plate) and spherical
waves (see appendix).  We also obtain three-body effects, in this case
on the equilibrium separation distance, but find that the magnitude
and sign of these effects depends strongly on the parameters of the
problem.

\section{Three-body Effects}
\label{sec:pairwise}

\begin{figure}[t!]
\includegraphics[width=1.0\columnwidth]{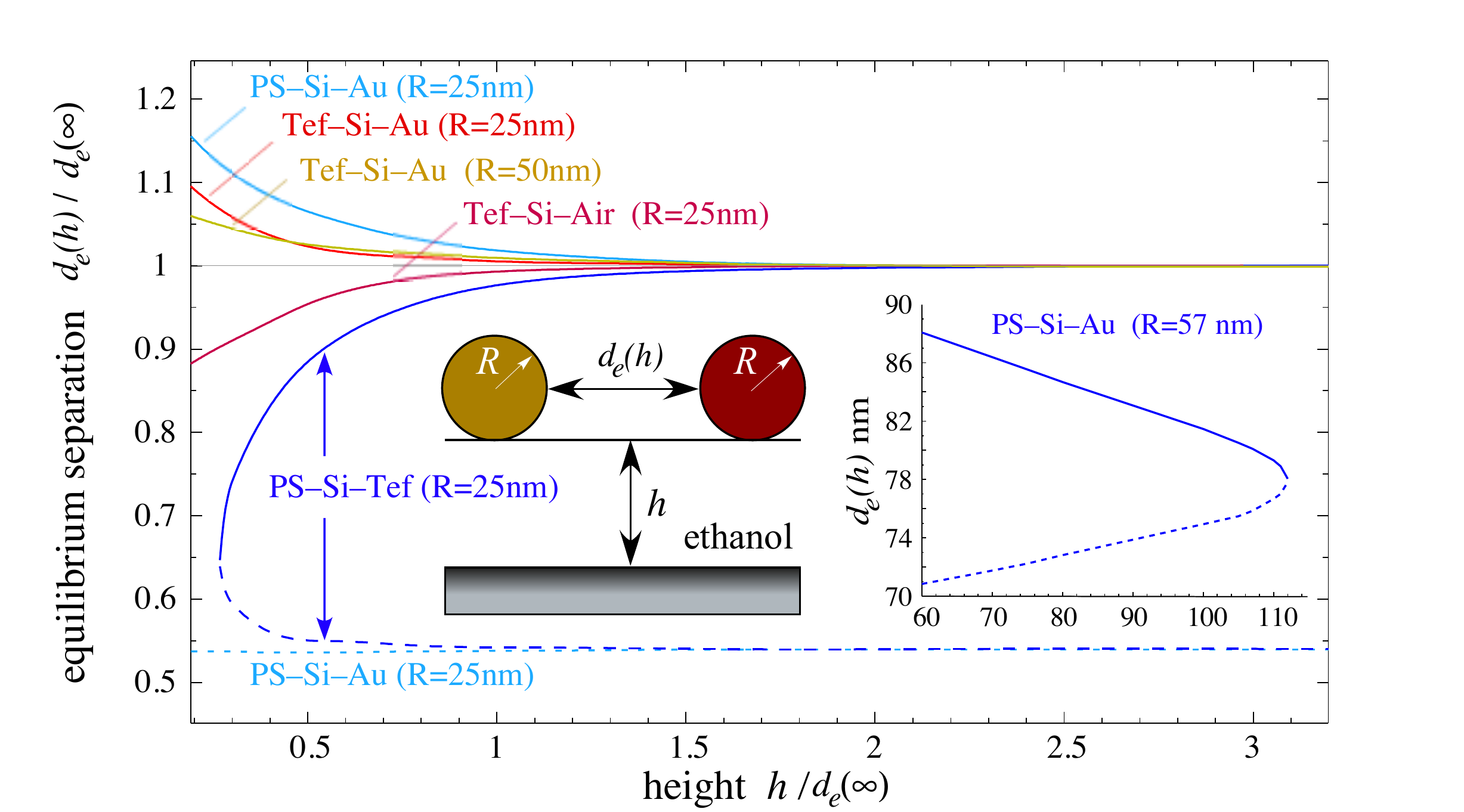}
\caption{Equilibrium separation $d_e(h) / d_e(\infty)$ between two
  $R=25$nm spheres suspended in ethanol as a function of their
  surface--surface separation $h$ from a plate (and normalized by the
  equilibrium separation for the case of two isolated spheres, i.e. $h
  = \infty$). $d_e$ is plotted for various material combinations,
  denoted by the designation sphere--sphere--plate, e.g. a PS and
  silicon sphere suspended above a gold plate is denoted as
  PS--Si--Au. Solid/dashed lines correspond to stable/unstable
  equilibria. (In the case of a gold plate, the spheres are chosen to
  have $R = 50$nm.)  The inset shows $d_e$ (in units of nm) for the
  case of two PS and silicon spheres ($R=57$nm) above a gold plate.}
\label{fig:fig2}
\end{figure}

To quantify the strength of three-body effects in the
sphere--sphere--plate system of \figref{fig1}, we begin by computing
how the zero--temperature equilibrium sphere--sphere separation $d$
varies as a function of the sphere-plate separation $h$ for two
equal-radius spheres, as plotted in \figref{fig2}.  To start with, we
consider very small spheres, with radius $R=25$~nm, for which the
three-body effects are substantial.  The separation $d_h$ at a given
$h$ is normalized by $d_\infty$ ($d$ as $h\to\infty$, i.e. in the
absence of the plate).  Several different material combinations are
shown (where $X$--$Y$--$Z$ denotes spheres of materials $X$ and $Y$
and a plate of material $Z$): polystyrene (PS), teflon (Tef), and
silicon (Si) spheres with gold (Au), teflon (Tef), and vacuum (air)
plates (the latter corresponding to a fluid-gas interface).  Depending
on the material combinations, we find that $d_h$ can either increase
or decrease by as much as $15\%$ as the plate is brought into
proximity with the spheres from $h=\infty$ to $h \approx R$. (We
expect even larger deviations when $h < R$, but small separations are
challenging for this computational method~\cite{Rahi09:PRD} and our
results for $h \geq R$ suffice here to characterize the general
influence of three-body effects.)

Interestingly, depending on the material combination, the $d_h$ can
either increase or decrease as a function of $h$: that is, the
proximity of the plate can either increase or decrease the effective
repulsion.  This is qualitatively similar to previous results for
vacuum-separated perfect-metal spheres/plates~\cite{Lopez09} in the
following sense.  Previously, the attractive interaction between a
sphere and a plate was in general found to enhance the attraction
between two identical spheres as the plate became closer
\cite{Lopez09}.(There are certain regimes, not present here, where the
attractive interaction decreases) Here, we observe that the
sphere--plate interaction changes the sphere--sphere interaction with
\emph{the same sign} as $h$ becomes smaller: if the sphere--plate
interaction is repulsive, the sphere--sphere interaction becomes more
repulsive (larger $d$), and vice-versa for an attractive sphere--plate
interaction.  Since the spheres are not identical, the three-body
effect is dominated by the sign of the stronger sphere--plate
interaction out of the two spheres.  Thus, examining the signs and
magnitudes of the pairwise interactions in all cases of \figref{fig2}
turns out to be sufficient to predict the sign of the three-body
interaction, although we have no proof that this is a general rule.
(In contrast, for non-spherical objects such as cylinders, there can
be competing three-body effects that make the sign more difficult to
predict, even in vacuum-separated geometries where all pairwise
interactions are attractive, which can even lead to a non-monotonic
effect~\cite{RahiRo07,Rodriguez07:PRL}.)

\Figref{fig2} also exhibits the interesting phenomenon of
bifurcations, in which stable equilibria (solid lines) and unstable
equilibria (dashed lines) appear/disappear at some critical $h$ for
certain materials and geometries, which is discussed in more detail in
\secref{bifurcations}.  As the sphere radius $R$ increases, all of
these three-body effects rapidly decrease, eventually entering an
additive regime in which three-body effects are negligible and in
which a parallel-plate/PFA approximation eventually becomes valid, as
described in \secref{additive}.

\subsection{Bifurcations}
\label{sec:bifurcations}

\begin{figure}[t!]
\includegraphics[width=1.0\columnwidth]{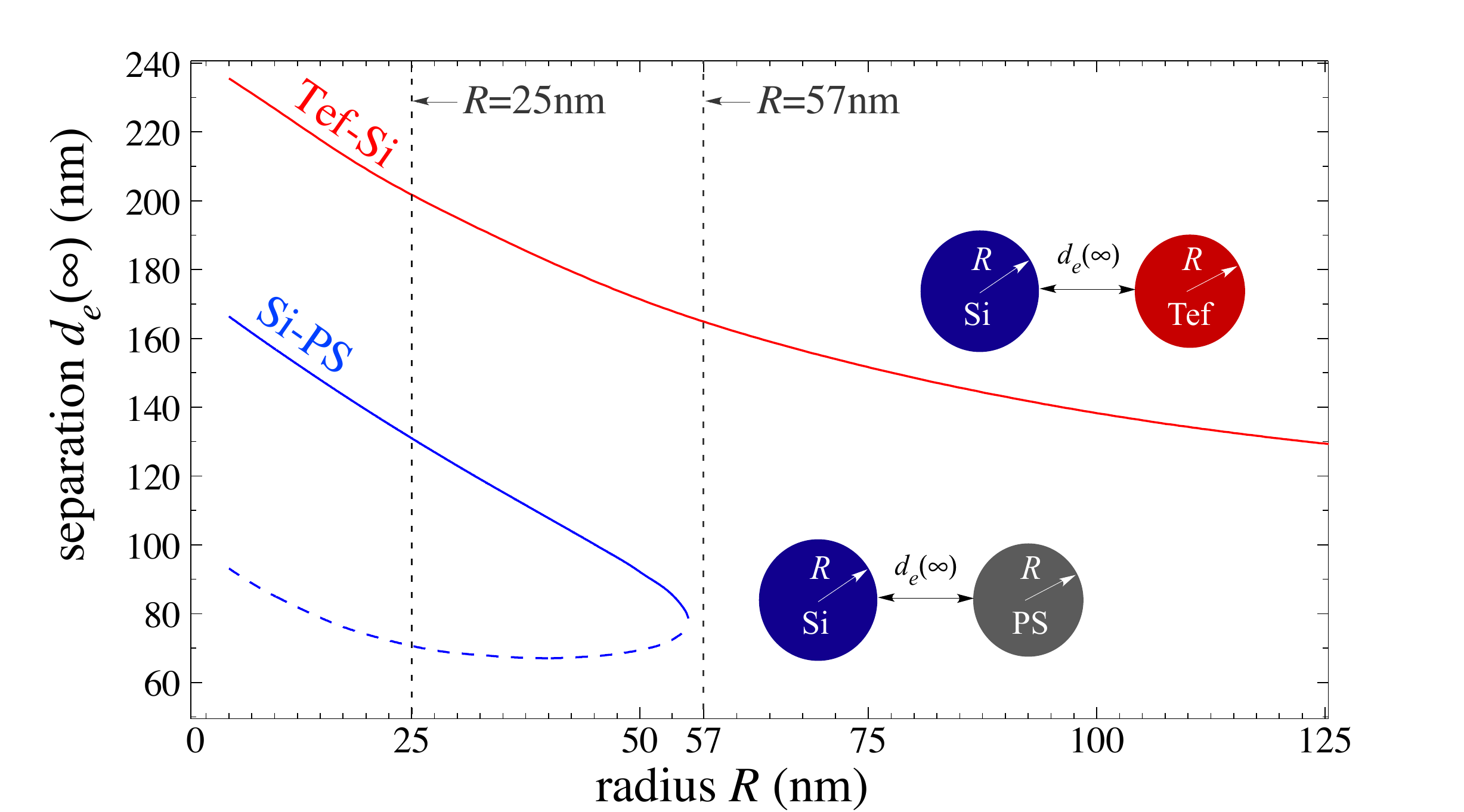}
\caption{Equilibrium separation $d_e(\infty)$ (units of nm) between a
  Si sphere and either a teflon(Tef) or polystyrene(PS) sphere
  immersed in ethanol as a function of their equivalent radii $R$.
  Solid/dashed lines denote stable/unstable equilibria.}
\label{fig:fig3}
\end{figure}

In the case of PS and Si spheres suspended above either a gold or
teflon plate, one can observe the emergence or disappearance of a
stable (solid) and unstable (dashed) pair of equilibria as $h$
decreases from $h=\infty$, respectively, as evidenced by the blue
curves in \figref{fig2} (teflon plate) and \figref{fig2}~(inset) (Au
plate).  This can be qualitatively explained by the fact that the
isolated sphere--sphere interactions exhibit a natural bifurcation for
sufficiently-large spheres, in conjunction with the fact that the
presence of the plate typically acts to either increase or decrease the
sphere--sphere interaction, depending on the sign of the dominant
sphere--plate interaction, as explained above.

In particular, \figref{fig3} shows the isolated Si--PS and Tef--Si
sphere--sphere equilibrium separation $d_e$ as a function of the
radius $R$ of the spheres.  As a consequence of its material
dispersion (similar to phenomena observed
in~\cite{RodriguezMc10:PRL}), the Si--PS combination exhibits a
bifurcation at $R \approx 55$~nm where the stable and unstable
equilibria, such that there is no equilibrium for larger $R$ (the
interaction is purely attractive).  The Tef--Si combination exhibits
no such bifurcation (even if we extend the plot to $R=300$~nm),
because it has no unstable equilibrium: the interaction is purely
repulsive for small separations and attractive for large separations.
Therefore, if the Si--PS radius is above or below the 55~nm
bifurcation, the presence of the plate can shift this bifurcation and
lead to a bifurcation as a function of $h$ as in \figref{fig2},
whereas no such bifurcation with $h$ appears for Tef--Si.

In the Si--PS--Au case of a gold plate with Si--PS spheres, the
sphere--plate interactions turns out to be primarily repulsive, which
should push the bifurcation in \figref{fig3} to the \emph{right}
(shrinking the attractive region) as $h$ decreases. Correspondingly,
if we choose a radius $R=57$~nm just to the right of isolated-sphere
bifurcation, then as $h$ decreases the Si--PS--Au combination should
push the bifurcation past $R=57$~nm leading to the creation of a
stable/unstable pair for small $h$, and precisely this behavior is
observed in the inset of \figref{fig2}.  Conversely, for the
Si--PS--Tef case of a teflon plate with Si--PS spheres, the
sphere--plate interaction is primarily attractive, and the opposite
behavior occurs: choosing a radius $R=25$~nm to the left of the
isolated-sphere bifurcation, decreasing $h$ increases the attraction
and moves the bifurcation to the \emph{left} in \figref{fig3},
eventually causing the disappearance of the stable/unstable
equilibrium at $R=25$~nm.  Correspondingly, for the Si--PS--Tef curve
in \figref{fig2}, we see the disappearance of a stable/unstable pair
for sufficiently small $h$.

\subsection{The additive regime}
\label{sec:additive}

In general, three-body effects can expected to disappear in various
regimes where key parameters of the interaction become small. First,
for large radii, where $h$ (the sphere--plate separation) and $d$ (the
sphere--sphere separation) become small compared to $R$, eventually
the Casimir interaction is dominated by nearest-surface interactions,
or the proximity-force approximation (PFA), in which the force can be
approximated by additive surface--surface ``parallel-plate'' forces
\cite{bordag01, Gies,Derjaguin}.  In order to damp the Brownian
fluctuations as described in the next section, we actually propose to
use much larger ($R>5\,\mu$m) spheres, and we quantify the accuracy of
PFA in this regime below.  Second, as $h$ becomes large compared to
$d$, the effect of the plate becomes negligible and three-body effects
disappear; this is apparent in \figref{fig2}, where $d_e \to
d_e(\infty)$ when $h \gg d_e$.  Third, in the limit where one of the
spheres is much smaller than the other sphere, then the smaller sphere
has a negligible effect on the sphere--plate interaction of the larger
sphere, and at least some of the three-body effect disappear as
described below.  In fact, we find that even for a situation in which
one sphere is only a few times smaller than the other, the three-body
effects tend to be negligible.  For the sphere-radius regime
considered in our previous work, we argue below that equal-height
suspension of the two spheres leads to a strong asymmetry in sphere
radii that tends to eliminate three-body effects.

\begin{figure}[t]
\includegraphics[width=1.0\columnwidth]{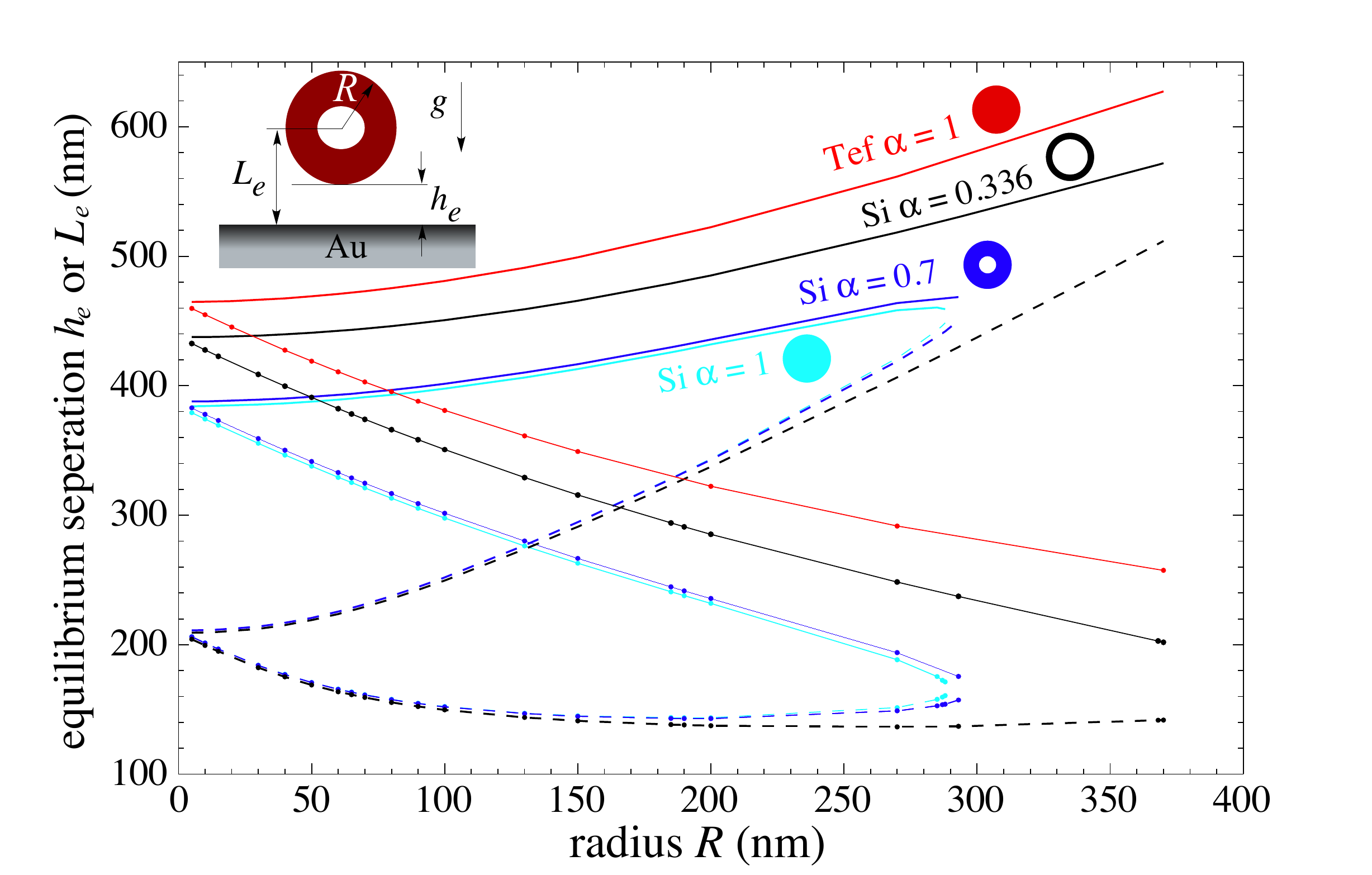}
\caption{Center--surface $L_e$ (solid lines) and surface--surface
  $h_e$ (dotted lines) equilibrium separation (units of nm) of a
  teflon (red) or Si hollowed sphere (shown on the inset) suspended in
  ethanol above a gold plate, as a function of radius $R$ (units of
  nm). The equilibria are plotted for different values of the
  fill-fraction $\alpha$, defined as the ratio of the spherical--shell
  thickness over the radius of the sphere. Solid/dashed lines
  correspond to stable/unstable equilibria.}
\label{fig:fig4}
\end{figure}

To begin with, let us consider sphere radii on the order of $10^2$~nm,
as in our previous work~\cite{RodriguezMc10:PRL}.  We wish to make a
bound dicluster, at some separation $d$, of two spheres (Si and
teflon) that are suspended above a gold substrate by Casimir repulsion
in balance with gravity.  Furthermore, suppose that we wish to suspend
both spheres at the same equilibrium height $h_e$, and therefore
choose the radii of the two spheres to equate their $h_e$ values.  In
\figref{fig4}, we plot $h_e$ as a function of radius $R$ for the
isolated sphere--plate geometries ($d\to\infty$).  For example, with
an Si sphere of radius $R=100$~nm, the (stable) equilibrium height is
$h_e = 298.17$~nm, whereas to obtain the same $h_e$ value for teflon
one needs a much larger teflon sphere of radius $R=217.2$~nm,
primarily because the Casimir repulsion is stronger for teflon.  If,
instead of a pairwise calculation, we perform an exact three-body
calculation of the $h_e$ values for these radii at the equilibrium
sphere--sphere separation $d_e = 92.8$~nm, we find that the $h_e$
values change by $< 1$\%.  Conversely, if we keep $h_e$ fixed and
compute the three-body change in $d_e$ (compared to $h\to\infty$),
again we find that the change is $<1$\%.  As mentioned above, the
small size of the Si sphere makes it unsurprising that the Si sphere
does not change the equilibrium $h_e$ of the much larger teflon
sphere.  Furthermore, the sensitivity of the sphere--sphere force
$F_d$ to the teflon $h$ is equal to the sensitivity of the teflon
sphere--plate force $F_h$ to $d$, thanks to the equivalence $\partial
F_d/\partial h = -\partial^2 U/\partial d \partial h = \partial F_h
/\partial d$ where $U$ is the energy.  Therefore, one would also not
expect the finite value of $h_e$ for the Si sphere to modify the
equilibrium $d_e$.  Size asymmetry alone, however, does not explain
why the finite $h_e$ of the \emph{teflon} sphere does not affect the
sphere-plate interactions of the Si sphere.  Even if the Si sphere
were of infinitesimal radius, the Casimir--Polder energy the Si sphere
would be determined by a Green's function at the Si
location~\cite{Lifshitz80}, and if the Si sphere is at comparable
distance $d_e \sim h_e$ from both the teflon plate and the sphere, one
would in general expect the response to a point-dipole source at the
Si location (the Green's function) to depend non-additively on the
teflon sphere and the plate even for an infinitesimal Si sphere.
However, in the present case we do not observe any non-additive effect
on the Si-sphere $h_e$, because the factor of three(approximately)
difference between $h_e$ and $d_e$ is already sufficient to eliminate
three-body effects (as in \figref{fig2}).

\Figref{fig4} also exhibits a bifurcation of stable (solid lines) and
unstable (dashed lines) equilibria that causes the stable $h$
equilibrium to vanish for Si at large radii.  In order to utilize
larger spheres to reduce the effects of Brownian motion in the next
section, one can consider instead a geometry of hollow air-filled
spherical shells with outer radius $R$ and shell thickness $\alpha R$
(so that $\alpha=1$ gives a solid sphere).  Such hollow microspheres
are readily fabricated with a variety of materials~\cite{Wilcox}.  As
\figref{fig4} shows, decreasing the shell thickness $\alpha$ pushes
the bifurcation to larger $R$, and also increase $h_e$ by making the
sphere more buoyant.  This modification allows us to consider
$R\approx10\,\mu$m in the next section, where PFA should be accurate.
For only $3\,\mu$m spheres and $500$~nm separations in fluids, we
previously found that the correction to PFA (which scales as $d/R$ to
lowest order~\cite{gies06:PFA,Maia-Neto,Mazzitelli}) was only about
15\%. For the three times larger radii and somewhat smaller
separations in the next section the corrections to PFA are typically
$< 5$\%, sufficient for our current purposes.

\section{Nonzero Temperature and Experiments}
\label{sec:temp}

In this section, we address a number of questions of consequence to an
experimental realization of the teflon/silicon two-sphere dicluster of
\figref{fig1}. In particular, we consider several ways in which a
nonzero temperature can disrupt the observation of stable
equilibria. A nonzero temperature will manifest itself in at least two
important ways. First, there will be a change in the Casimir force
between the objects due to the presence of real (non-virtual) photons
in the system. Second, the inclusion of nonzero temperature will cause
the spheres to experience Brownian motion arising from the thermal
agitations in the fluid~\cite{Risken}. We consider the influence of
both of these effects on the observability of stable particle clusters
and suspensions.

At zero temperature, the Casimir force $F$ is determined by an
integral $F = \int_0^\infty d\xi f(\xi)$ of a complicated integrand
$f(\xi)$ evaluated at imaginary frequencies
$\xi$~\cite{Lifshitz80}. At $T > 0$, the integral is replaced by a
finite sum over Matsubara frequencies $\omega_n =2\pi n k T/\hbar$,
arising from the poles of the $\coth$ photon distribution along the
imaginary frequency axis ~\cite{schwinger78, Bordag}, leading to a force
$F_T$ given by:
\begin{equation}
  F_T = \frac{2\pi k T}{\hbar} \left[\frac{f(0+)}{2} + \sum_0^\infty
    f\left(\frac{2\pi kT}{\hbar}n\right)\right],
\end{equation}
which is exactly a trapezoidal-rule approximation to the
zero-temperature force with a discretization error determined by the
Matsubara wavelength $\lambda_T = 2\pi c / \xi_T = \hbar /
kT$~\cite{boyd01:book}. Because the integrand $f(\xi)$ is smooth and
typically varies on a scale much slower than $1/\lambda_T$, where
$\lambda_T = 7.6\,\mu$m at room temperature $T=300$~K, the finite-$T$
correction to the zero-temperature Casimir force is often
negligible~\cite{milton04}.  However, in fluids, as is the case here, larger
temperature effects have been obtained ~\cite{RodriguezW10:PRL} by
dispersion-induced oscillations in $f(\xi)$, and so we must check our
previous zero-temperature predictions against finite-$T$ calculations.
For the Tef--Si--Substrate case considered here, we find that $T>0$
corrections to the $T=0$ forces are no more than 2\% over the entire
range of separations considered here, and hence they can be neglected.

The presence of Brownian motion proves a much more difficult
experimental complication to overcome. First, Brownian motion will
lead to random fluctuations in the position of the spheres, making it
hard to measure their stable separations in an experiment
~\cite{Risken}. Second, and more importantly, sufficiently large
fluctuations can drive the Si sphere to ``tunnel'' past its unstable
equilibrium position with the gold plate, leading to
stiction~\cite{Risken} since the Si--Au interaction is purely
attractive for small separations. The remainder of this section will
revolve around the question of how and whether one can overcome both
of these difficulties to observe suspension in experiments. In
particular, we consider observation of the average separation of the
spheres over a sufficiently long time, but not so long that stiction
occurs, and analyze the separation statistics and the stiction
timescale.  First, however, we describe how the parameters are chosen
so that Brownian fluctuations are not so severe.

\begin{figure}[t!]
\includegraphics[width= 0.5\columnwidth]{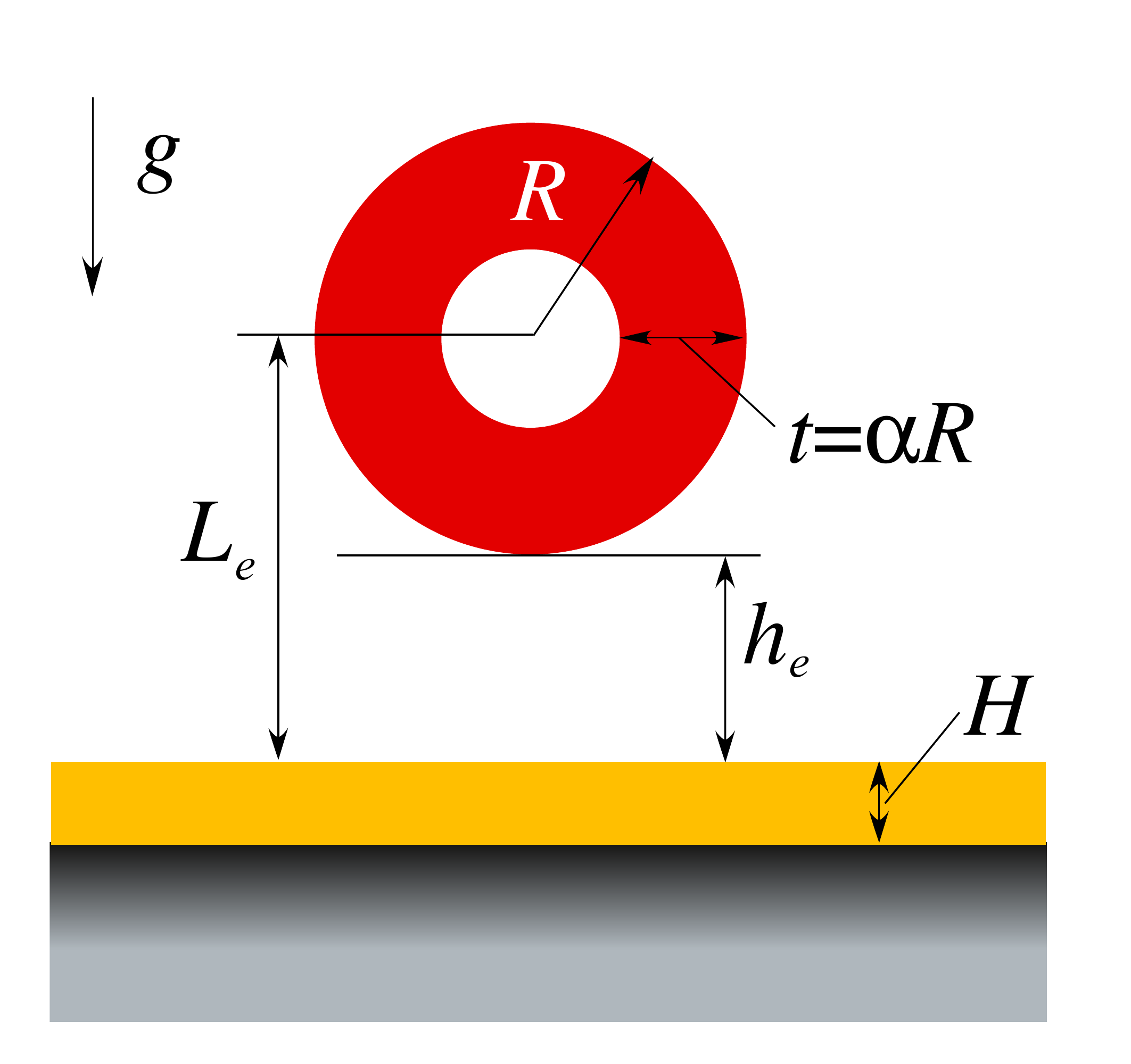}
\caption{Geometry of a hollow air--filled core sphere suspended above
  a layered plate with layer thickness $H$.  Explicitly shown is the
  thickness dimension as a function of $\alpha$}
\label{fig:fig5}
\end{figure}

The sphere geometry that we consider is depicted in \figref{fig5}: a
hollow spherical shell suspended by a surface--surface separation $h$
above a layered substrate, consisting of a thin indium tin oxide (ITO)
film of thickness $H$ deposited on a gold substrate, where the purpose
of the ITO layer is to eliminate the Si-sphere instability as
explained below. The thickness of the shell is denoted as $t=\alpha
R$, where $\alpha$ is a convenient fill-fraction parameter. We
consider hollow spheres in order to increase $R$ and thereby reduce
Brownian fluctuations. In particular, both the Brownian fluctuations
and the probability of stiction in the case of the silicon sphere are
reduced by increasing the strength of the Casimir force, which can be
achieved by increasing $R$ since the Casimir force scales roughly with
surface area, and below we consider radii from~1 to~10~$\mu$m.  In
this regime, as quantified in the previous section, a simple PFA
approximation is sufficient to accurately compute the forces and
separations.  However, because the gravitational force scales as
$R^3$, for large $R$ the gravitational force will overcome the Casimir
force and push the Si sphere past its unstable equilibrium into
stiction.  In order to reduce the gravitational force while keeping
the surface area fixed, we propose using a hollow Si sphere. We find
that in addition to hollowing the spheres, it is also beneficial to
deposit a thin ITO film,o n top of the gold substrate (the
permittivity of ITO is modeled via an empirical Drude model with
plasma frequency $\omega_p=1.46\times 10^{15}$~rad/s and decay rate
$\gamma = 1.53\times 10^{14}$~rad/s).  The ITO layer acts to decrease
the equilibria separations and therefore increase the Casimir
interactions between the spheres and the substrate. However, because
the Casimir force between teflon/silicon and ITO is
attractive/repulsive at small separations, respectively, increasing
$H$ pushes the silicon-substrate's unstable equilibrium to smaller
separations while \emph{introducing} a teflon-substrate unstable
equilibrium that gets pushed to larger separations. In what follows,
we find that $H$ from 14--30~nm is sufficient to obtain experimentally
feasible suspensions, although here we only consider the case of
$H=15$~nm.

\begin{figure}[t!]
\includegraphics[width=1.0\columnwidth]{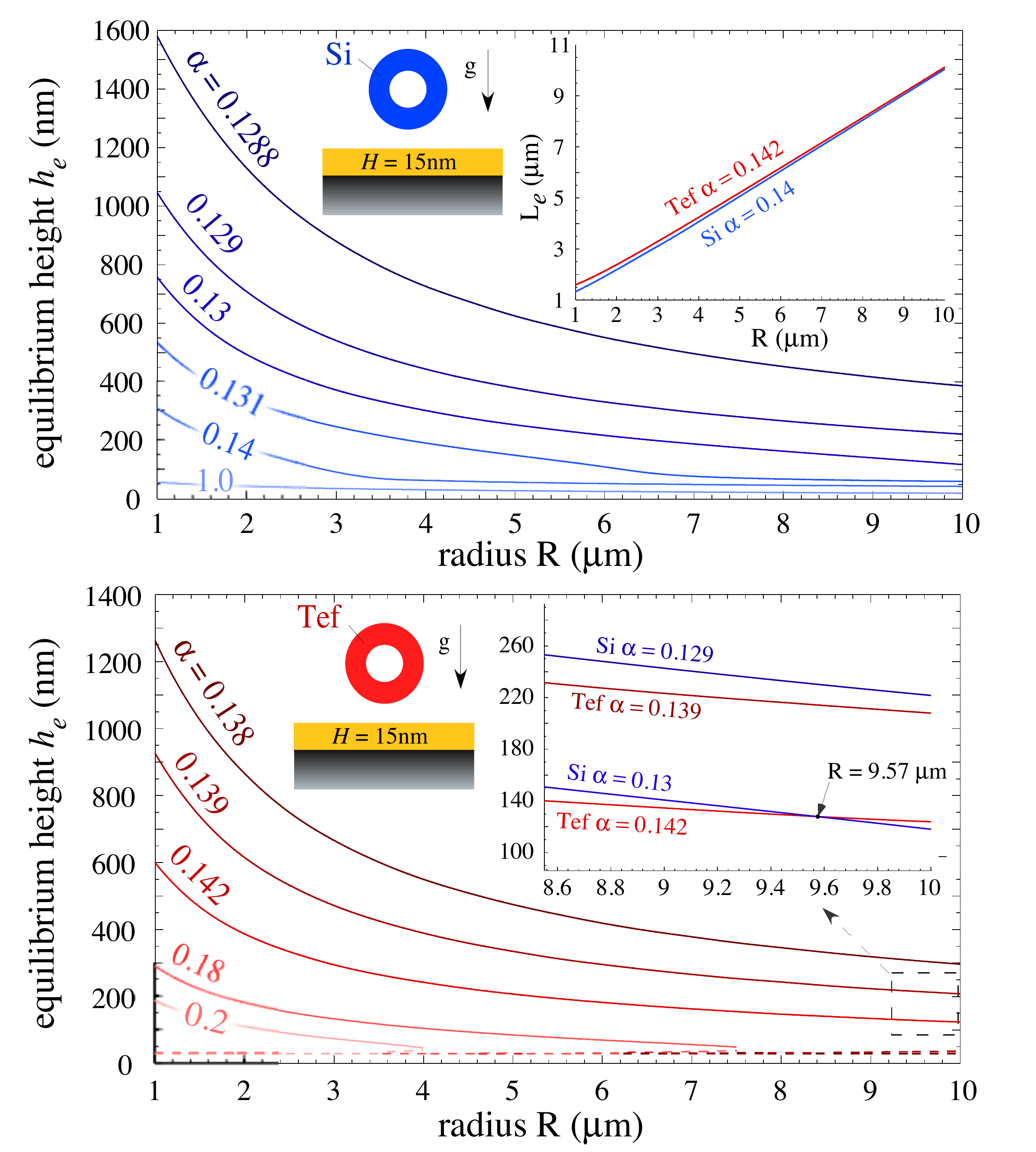}
\caption{Surface--surface equilibrium height $h_e$ (units of nm) for
  the hollowed--sphere geometry of \figref{fig5}, consisting of either
  a Si (top) or teflon (bottom) hollowed sphere (fill-fraction
  $\alpha$) suspended in ethanol above a $H = 15$~nm ITO layered gold
  plate, as a function of sphere radius $R$ (in units of
  $\mu$m). Solid/dashed lines correspond to stable/unstable
  equilibria. $h_e$ is plotted for different values of $\alpha$,
  denoted in the figure. The top inset plots the center--surface
  separation $L_e$ (in units of nm) as a function of $R$ of a hollowed
  teflon/Si (red/blue lines) sphere suspended again above a gold
  plate, for $\alpha = 0.14/0.142$. The lower inset shows $h_e$ for
  both teflon and Si spheres for $R \in [8.6,10]\mu$m.}
\label{fig:fig6}
\end{figure}

The effect of hollowing the spheres is shown in the top panel of
\figref{fig6}: smaller $\alpha$ values push the stable/unstable
bifurcation of teflon out to larger $R$.  Hollowing the silicon sphere
is not necessary because silicon has no unstable equilibrium (in this
configuration it is repulsive down to zero separation).  However, as
shown in the bottom panel of \figref{fig6}, hollowing the silicon
sphere does change its $h_e$ at a given $R$.  For example, one can
choose a Tef $\alpha=0.142$ and a Si ($\alpha=0.14$) to
obtain the same equilibrium surface-to-center height $L_e$ over a wide
range of sphere radii, as shown in upper-right inset of \figref{fig6}.
Alternatively, one can choose a hollow teflon sphere to match the
equilibrium surface-surface separations $h_e$ for equal sphere radii,
as shown in the lower-right inset of \figref{fig6}.

\begin{figure}[t!]
\includegraphics[width=1.0\columnwidth]{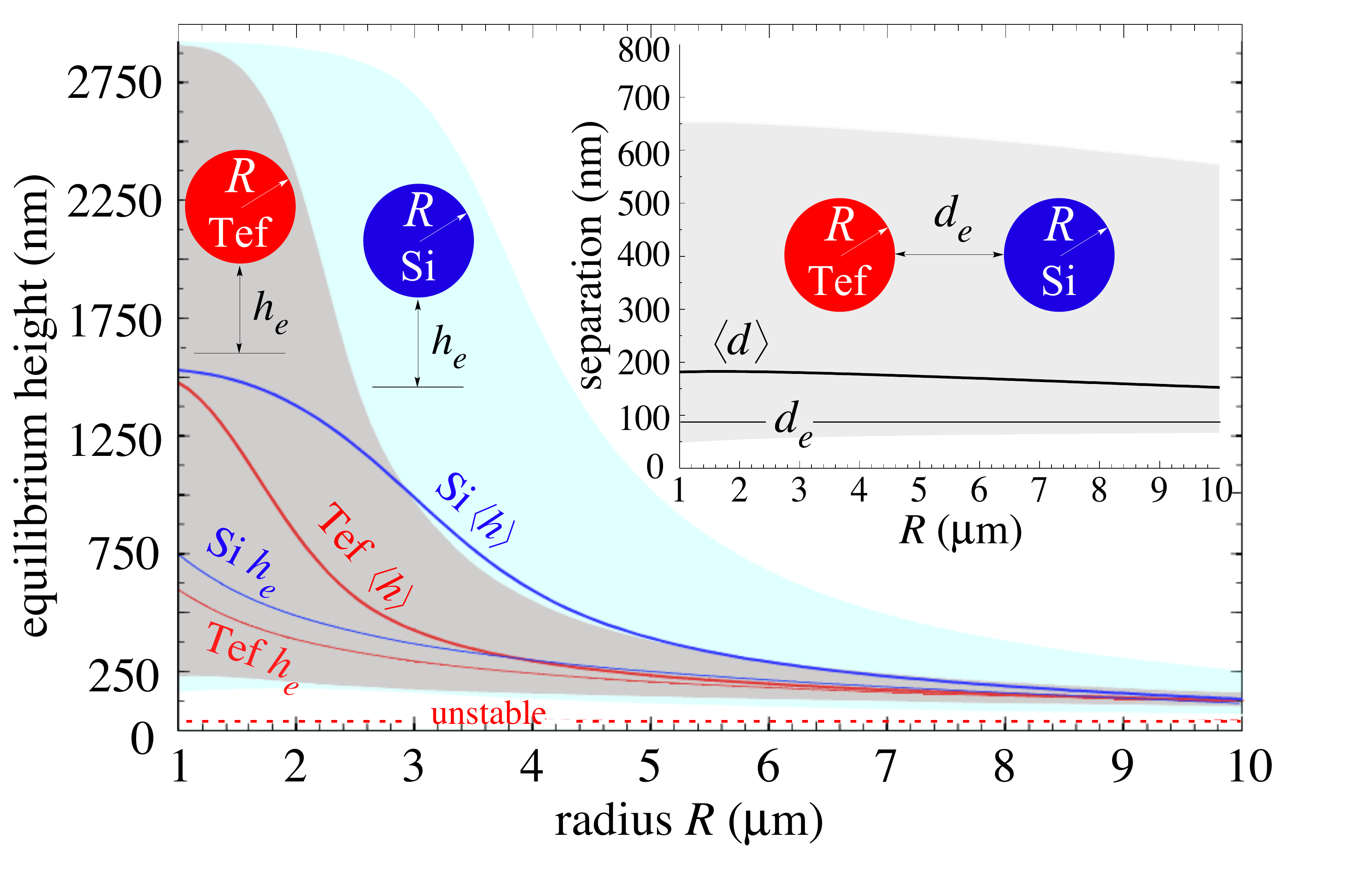}
\caption{Average $\langle h \rangle$ (thick lines) and equilibrium
  $h_e$ (thin lines) height (in units of nm) of a hollowed teflon/Si
  (blue/red lines) sphere suspended above a $H = 15$~nm ITO layered
  gold plate, for $\alpha = 0.142/0.13$, as a function of sphere
  radius $R$ (in units of $\mu$m). Solid/dashed lines correspond to
  stable/unstable equilibria. The red/blue shaded regions indicate
  positions where the teflon/Si spheres are found with $95\%$
  probability.  The inset shows $\langle d \rangle $ (thick line) and
  $d_e$ (thin line) separations as a function of their radius for two
  equal radii teflon Si spheres.  The gray shaded region indicate the
  separations which the teflon and Si spheres are found with $95\%$
  probability.}
\label{fig:fig7}
\end{figure}

\subsection{Statistics of Brownian motion}

As mentioned above, Brownian motion will disturb the spheres by
causing them to move randomly about their stable equilibrium
positions, and this can cause the Si sphere to move past its unstable
equilibrium point, inducing it to stick top the plate. To quantify the
range of motion of both spheres about their equilibria, we consider
the statistical properties of their fluctuations. In particular, we
consider the average plate--sphere separations $\langle h \rangle_T$
and average sphere--sphere separations $\langle d \rangle_T$ near room
temperature ($T=300$~K), determined by an ensemble average over a
Boltzman distribution. For example, $\langle h \rangle_T$ is given
by:
\begin{equation}
  \langle h \rangle_T = \dfrac{\int_0^\infty dz\, z
    \exp\left(U(z)/kT\right)}{\int_0^\infty dz\,
    \exp\left(U(z)/kT\right)},
\end{equation}
where $U(z)$ is the total energy (gravity included) of the
sphere--plate system at a surface--surface height $z$. (A similar
expression yields $\langle d \rangle_T$).  In the case of teflon, the
short-range attraction means that the suspension is only metastable
under fluctuations; here, we only average over separations prior to
stiction by restriction $z$ to be $\geq$ the unstable equilibrium, and
consider the stiction timescale separately below.  In addition to the
average equilibrium separations, we are also interested in quantifying
the extent of the fluctuations of the spheres, which we do here by
computing the 95\% confidence interval $\{\sigma_{-},\sigma_{+}\}$,
defined as the spatial region over which the sphere is found with 95\%
probability around the equilibria, where $\sigma_{\pm}$ denotes the
lower/upper bound of that interval.  These results are shown in
\figref{fig7} for $h$, with $d$ shown in the inset, in which shaded
regions indicate the confidence intervals, as a function of $R$ where
$\alpha$ is chosen to yield approximately equal $h_e$ ($\alpha =
0.142$ for teflon and $\alpha=0.13$ for Si).  (Note that the
horizontal separation $\langle d \rangle$ is a purely Casimir
interaction and the difference here from $\alpha=1$ is negligible in
the PFA regime.)  As predicted above, the Brownian fluctuations of the
spheres vanish as $R \to \infty$ and are dramatically suppressed for
$R \gtrsim 5\,\mu$m, where one finds $\langle h \rangle \approx
h_e$. In addition, we find that the teflon sphere can safely avoid the
unstable equilibrium and stiction in the sense that the unstable
equilibrium is far outside the confidence interval; the timescale of
the stiction process is quantified below.  The asymmetrical nature of
the confidence interval results from the fact that the Casimir energy
decreases as a function of $z$, and as a consequence the Brownian
excursions favor the $+z$ direction.  The fluctuations in $\langle d
\rangle$ are substantially larger than those in $\langle h \rangle$
(nor is there any obvious reason why they should be comparable, given
that the nature of the sphere--sphere equilibrium is completely
different from the sphere--plate equilibrium), making the precise
value of $d_e$ potentially harder to observe.

\begin{figure}[t]
\includegraphics[width=1.0\columnwidth]{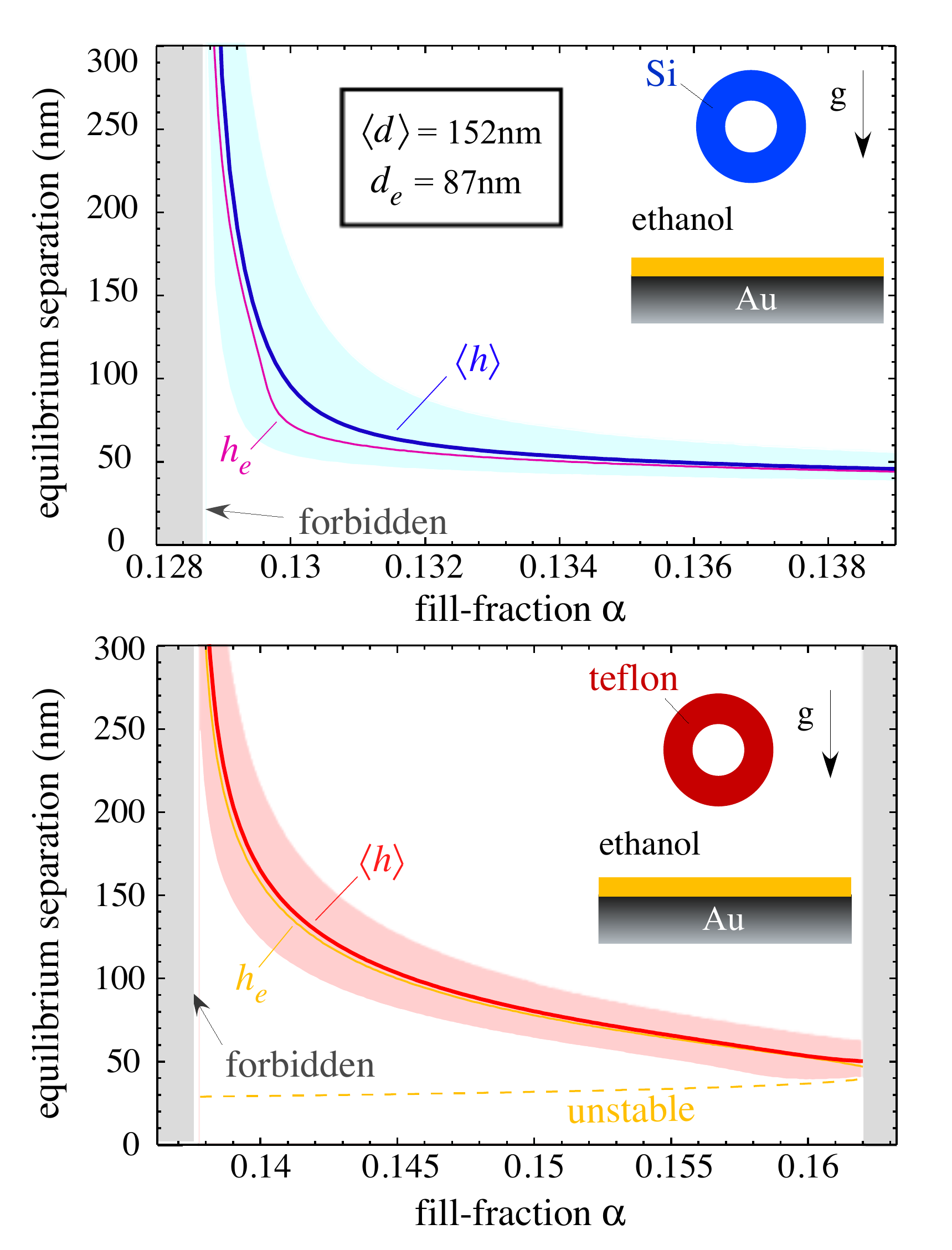}
\caption{Average $ \langle h \rangle$ (thick line) and equilibrium
  $h_e$ (thin line) height (in units of nm) of a hollowed Si/teflon
  sphere (top/bottom) of radii $R = 10$/$9.915\mu$m suspended in
  ethanol above a $H = 15$~nm ITO layered gold plate, as a function of
  fill-fraction $\alpha$ (indicated in \figref{fig5}).  Shaded regions
  indicate $h$ positions where the Si/teflon spheres are found with
  $95\%$ probability. Solid/dashed lines indicate stable/unstable
  equilibria. For reference, we state the equilibrium $d_e$ and
  average $\langle d \rangle$ horizontal seperations between $R =
  10$/$9.915\mu$m Si/Tef spheres in the top figure.}
\label{fig:fig9}
\end{figure}

Instead of considering the Brownian statistics as a function of $R$,
we can instead consider the statistics as a function of $\alpha$ for
fixed radii $\approx 10\mu$m (chosen to obtain nearly equal
sphere-center heights $L_e$), as shown in \figref{fig9}.  One key
point is that there is a minimum allowed $\alpha$: if $\alpha$ is too
small, the buoyant force (assuming an air-filled hollow sphere) will
eventually become positive and the sphere will float, although this
limitation is removed if one could infiltrate the hollow sphere with
the fluid.  For the teflon sphere, there is also an upper limit to
$\alpha$ for a given $R$ to avoid stiction as discussed previously.

\subsection{Stiction and tunnelling rates}

As mentioned above, the stable equilibrium for the teflon sphere is
actually only metastable---because the Casimir force is attractive for
small separations, given a sufficiently long observation time $\tau$
the sphere will ``tunnel'' (via Brownian fluctuations) past the energy
barrier $\Delta$ posed by the unstable equilibrium, and stick to the
plate (stiction).  Given the energy barrier, the temperature $T$, and
the viscous drag on the particle, we can apply standard
methods~~\cite{Chandrekasar,Risken,Melnikov19911} to compute the
timescale for stiction.  This calculation, which is described in
detail below, shows that for various values of the fill factor
$\alpha$ the expected time $\tau$ to stiction (which increases
exponentially with $\Delta/kT$) can vary dramatically, but can easily
be made on the order of years.

\begin{figure}[t]
\includegraphics[width=1.0\columnwidth]{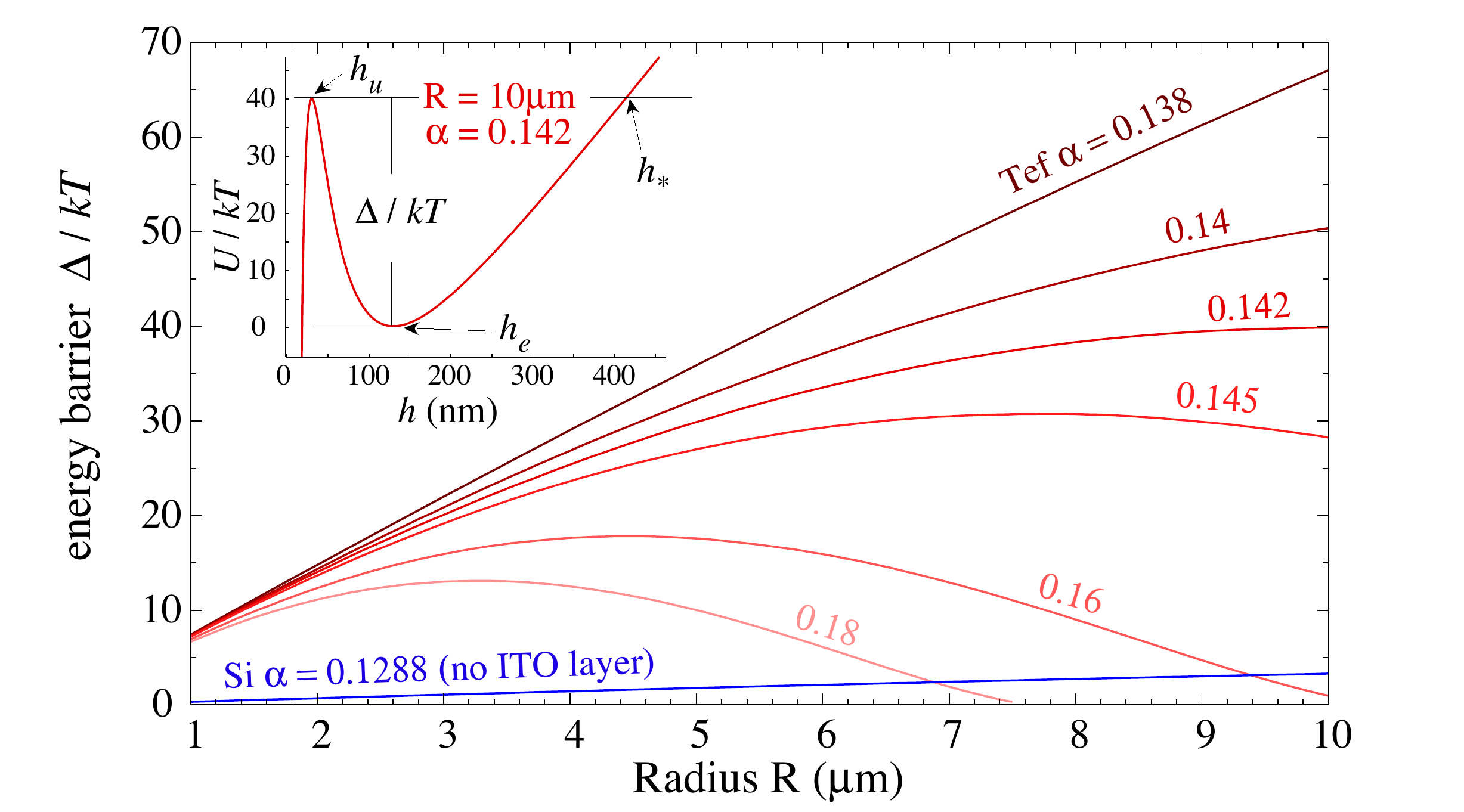}
\caption{Energy barrier $\Delta / kT$ of a hollowed teflon sphere
  suspended in ethanol above a $H = 15$~nm ITO layered gold plate at
  $T=300$~K, as a function of sphere radius $R$ (in units of $\mu$m)
  and for different values of fill-fraction $\alpha$. The inset shows
  the energy landscape $U / k T$ as a function of the surface--surface
  height $h$ (units of nm) for a teflon sphere of radius $R=10\mu m$
  with a fill fraction of $\alpha = 0.142$. }
\label{fig:fig8}
\end{figure}

The energy barrier $\Delta/kT$ is plotted versus the teflon sphere
radius $R$ for various $\alpha$ in \figref{fig8}, and can easily be
made $> 10$ to obtain a very long metastable lifetime.  As we
discussed earlier, the $\Delta$ increases with $R$ at first because
this increases the Casimir force, but has a maximum at some $R$ where
gravity begins to dominate.  Decreasing $\alpha$ decreases the
gravitational force and therefore increases both the maximum $\Delta$
and the corresponding $R$.  A typical energy landscape $U(z)$ is shown
in the inset, exhibiting a local minimum at a height $h_e$ and an
unstable equilibrium (maximum) at $h_u$.  Also noted on the inset is
the ``tunneling'' height $h_* > h_e$ at which $U(h_*) =
U(h_u)$. \Figref{fig8} also shows the energy barrier $\Delta/kT$ of a
silicon sphere ($\alpha = 0.1288 \approx \alpha_c$, $R=10~\mu$m) in
the absence of the ITO layer ($H = 0$) to be significantly smaller
than that of teflon. Of course $\Delta / kT$ in this case could be
made larger merely by choosing $\alpha \approx \alpha_c$, but we find
(below) that achieving experimentally realizable lifetimes severely
limits the range of realizable $\alpha$, i.e. requires that the Si
thickness be known to within a few nanometers.

Because $\Delta \gg kT$, the lifetime~$\tau$ of a Brownian particle
trapped around a local minimum of a potential $U(z)$ can be
approximated by~\cite{Melnikov19911}:
\begin{equation}
  \tau = e^{\Delta/kT}
   \left[\left(1 +
    \frac{\gamma}{4\omega^2}\right)^{1/2} -
    \frac{\gamma}{2\omega}\right]^{-1} 
    \frac{2\pi}{\Omega}
   \zeta\left(\frac{\gamma S}{kT}\right),
\label{eq:lifetime}
\end{equation}
where $\gamma$ is the viscous drag coefficient (drag force =
$-\gamma\,\mathrm{velocity}$), $\omega$ and $\Omega$ characterize the
curvature of $U(z)$ at the energy maximum and minimum respectively [as
  defined in \eqref{omega}], $\zeta(\delta)$ is a transcendental
function defined in \eqref{zeta}, and $S$ is an integral of the
potential barrier defined by \eqref{S}.  Let $m$ be the mass of the
sphere.  The drag coefficient for a sphere of radius $R$ in a fluid
with viscosity $\eta$ is $\gamma = 6 \pi R \eta /
m$~\cite{Landau:fluid}, where a typical viscosity is $\eta \approx
1.17 \pm 0.06$~mPas for ethanol~\cite{rhodata}.  The other quantities
are given by:
\begin{align}
  S &= 2 \int_{h_{u}}^{h_{c}} \, dz \sqrt{-2 m U(z)} \label{eq:S} \\
  \omega&=\sqrt{\frac{U''(h_u)}{m}} ; \Omega=\sqrt{\frac{U''(h_e)}{m}}
 \label{eq:omega} \\ \zeta(\delta) &= \exp\left[-\frac{2}{\pi} \int_0^{\pi/2} \,dz
    \ln\left(1 - e^{-\delta / 4 \cos^2 z}\right)\right]. \label{eq:zeta}
\end{align}
Combining these formulas and choosing different values of $R$ and
$\alpha$ to obtain different barriers $\Delta$ and landscapes $U(z)$
as in \figref{fig8}, the lifetime $\tau$ can be designed to take on a
wide range of values. The exponential dependence on $\Delta$ means
that $\tau$ rapidly transitions from very short to very long as
$\alpha$ changes, but can easily be made large.  For example, with $R=
8.5\mu$m and $\alpha < 0.15$, one obtains $\tau > 40$~days.
(Conversely, for sufficiently large $\alpha$ one could design
experiments where stiction occurs on an arbitrarily fast timescale,
but in this $\Delta \sim kT$ regime the approximations of
\eqref{lifetime} are no longer valid.)

Strictly speaking, this is a conservative estimate of the timescale
because the drag coefficient $\gamma$ for a sphere above a plate is
larger than that of an isolated sphere.  As the sphere approaches the
plate, the drag is dominated by the ``lubrication'' problem of the
fluid squeezed between the sphere and the plate, and the drag
increases dramatically~\cite{Hamrock}.

\section{Conclusion}
\label{sec:conc}

Even including the thermal motion of the particles and the finite
lifetime of metastable suspensions, the stable suspension and
separation of particle diclusters appears to be experimentally
feasible. In the experimentally relevant regimes, these effects
consist primarily of pairwise sphere--sphere and sphere--plate
interactions; while three-body effects become significant for smaller
spheres, the increased Brownian fluctuations for small spheres makes
such an experiment challenging.  Although the systems considered here
consisted of silicon and teflon spheres above layered substrate in
ethanol, many other materials combinations could potentially be
explored to modify these phenomena, including multi-material sphere
systems such as multi-layer spheres or patterned substrates that could
exhibit unusual effective dispersion phenomena.  Although we
considered hollow (air core) spheres, one could also use fluid-filled
spheres or similar modifications in order to modify the effect of
gravity.  Alternatively, one could use non-spherical geometries such
as disks, which have a both surface area and volume proportional to
$R^2$ so that gravity does not dominate asymptotically.  We have
recently demonstrated computational methods capable of accurate
modeling of such geometries, and find that the additional rotational
degrees of freedom can lead to additional phenomena such as
transitions in the stable orientation with
separation~\cite{Reid:arxiv}.  In general, the possibility of both
repulsion and stable equilibria in fluids (whereas the latter are not
possible in vacuum~\cite{Rahi10:PRL}but do exist in the critical
casimir fluids~\cite{Trondle, Mohry:PRE}) opens the possibility of a rich
and currently little explored territory for Casimir physics, and it is
likely that many effects remain to be discovered.

\section*{Appendix}

In what follows, we write down an expression for the Casimir energy of
of the system in~\figref{fig1}, in terms of the scattering and
translation matrices of the individual objects (spheres and plates) of
the geometry. A similar expression was derived in~\cite{Lopez09} in
the case of perfect-metal vacuum-separated objects, for which an
additional simplification, based on the method of images, was
possible~\cite{Brown:1969}. Here, we consider the more general case of
fluid-separated dielectric objects.

The starting point of the Casimir-energy expression is the well-known
scattering-matrix formalism, derived in~\cite{Emig07,Rahi09:PRD}, in
which the Casimir energy $U$ between an arbitrary set of objects can
be written as:
\begin{equation}
U=\frac{\hbar c}{2 \pi}
\displaystyle\int_{0}^{\infty}d\kappa
\log{\det{\mathbb{M}\mathbb{M}_{\infty}^{-1}}},
\label{eq:caseq}
\end{equation}
where $\mathbb{M}_{\infty}^{-1} = \mathrm{diag}(\mathbb{F}_{1} ,
\mathbb{F}_{2},...)$ and the matrix $\mathbb{M}$ is given by:
\begin{equation}
\mathbb{M} =\left( \begin{array}{cccc} \mathbb{F}_{1}^{-1} &
  \mathbb{X}^{12} & \mathbb{X}^{13} & ...\\ \mathbb{X}^{21} &
  \mathbb{F}_2^{-1} & \mathbb{X}^{23} & ...\\ ... & ... & ... & ...\\
\end{array}
\right),
\label{eq:MMinf}
\end{equation}
where $\mathbb{F}_i(\kappa)$ is the matrix of inside/outside
scattering amplitudes of the $i$th object, and $\mathbb{X}^{ij}$ the
translation matrix that relates the scattering matrix of the $i$th and
$j$th objects, as described in~\cite{Rahi09:PRD}. Here, the plate is
labeled by the index $i=1$ whereas the left and right spheres are
labeled as $i=2$ and $i=3$, respectively.

For computational convenience, the determinant in \eqref{caseq} can be
re-expressed in terms of standard operations on the block matrices
composing $\mathbb{M}$, and in this case we find that:
\begin{multline}
  \det \mathbb{M}\mathbb{M}_{\infty} =
  \det\left(\mathcal{I}-\mathcal{N}^{(1)}\right) 
  \det\left(\mathcal{I}-\mathcal{N}^{(2)}\right) \nonumber \\ \times
  \det\left(\mathcal{I}-\left(\mathbb{I}-\mathcal{N}^{(2)}\right)^{-1}
  \mathcal{A}\left(\mathcal{I}-\mathcal{N}^{(1)}\right)^{-1}\mathcal{B}\right),
\label{eq:detex}
\end{multline}
where
\begin{eqnarray}
\mathcal{N}^{(2)} =
\mathbb{F}_3\bb{X}^{31}\bb{F}_1\bb{X}^{13}, \,\,\, 
\mathcal{A}=F_{3}X^{32}-F_{3}X^{31}F_{1}X^{12};
\nonumber \\ 
\mathcal{B}=F_{2}X^{23}-F_{2}X^{21}F_{1}X^{13}, \,\,\, 
\mathcal{N}^{(1)}=F_{2}X^{21}F_{1}X^{12},
\label{eq:matrices}
\end{eqnarray}
where $(\mathcal{I}-\mathcal{N}^{(1)})$ and
$(\mathcal{I}-\mathcal{N}^{(2)})$ yield the individual interaction energies
of the left and right spheres with the plate, respectively. Because of
the logarithm in~\eqref{caseq}, it is possible to re-express the
energy as:
\begin{equation}
  U = \mathcal{E}_{1}(h_1)+\mathcal{E}_{2}(h_2)+\mathcal{E}_{int}(h_1,h_2, d),
\label{eq:Newsum}
\end{equation}
where,
\begin{eqnarray}
\mathcal{E}_{1}(h_1) = \frac{\hbar c}{2 \pi}
\displaystyle\int_{0}^{\infty}d\kappa \log \det
\left(\mathcal{I}-\mathcal{N}^{(1)}\right) \nonumber \\
\mathcal{E}_{2}(h_2) = \frac{\hbar c}{2 \pi}
\displaystyle\int_{0}^{\infty}d\kappa \log \det
\left(\mathcal{I}-\mathcal{N}^{(2)}\right),
\label{eq:twobodyens}
\end{eqnarray}
are the individual interaction energies of the left (1) and right (2)
spheres above a plate, in the absence of the other sphere, and
$\mathcal{E}_{int}(h_1,h_2,d)$ is a three-body interaction term given
by:
\begin{align}
\mathcal{E}_{int}=\frac{\hbar c}{2 \pi}\displaystyle \int d\kappa \log
\det&\left[\mathcal{I}-\left(\mathbb{I}-\mathcal{N}^{(2)}\right)^{-1}
  \right. \nonumber \\ &\times
  \left. \mathcal{A}\left(\mathcal{I}-\mathcal{N}^{(1)}\right)^{-1}\mathcal{B}\right],
\label{eq:threeterm}
\end{align}

Finally, for completeness, we write down simplified expressions for
the intermediate matrices $\mathcal{N}^{(i)}$, $\mathcal{A}$ and
$\mathcal{B}$, in terms of appropriate and rapidly-converging multipole
and Fourier basis, as explained in~\cite{Rahi09:PRD}.  The expression
for $\mathcal{E}_{1,2}$ was derived in ~\cite {Rahi09:PRD} and thus
here we can simply quote the result for the matrices
$\mathcal{N}^{(1)}$ and $\mathcal{N}^{(2)}$. In
particular,~\cite{Rahi09:PRD} expresses the matrices in terms of a
spherical multipole basis, indexed by the quantum numbers $l$, $m$,
and $P$, corresponding to angular momentum, azithmutal angular
momentum, and polarization [TE ($P=E$) or TM ($P=M$)]. The matrices
$\mathcal{N}^{(i)}$ are given by:
\begin{align}
\mathcal{N}^{(j)}_{lmP,l'm'P'} &= \delta_{m,m'}\mathcal{F}^{ee(j)}_{lmP,lmP} \nonumber \\
& \times\displaystyle\int_{0}^{\infty} \frac{k_{\bot}dk_{\bot}}{2 \pi}
\frac{e^{-2h_{j}\sqrt{\vec{k}_{\bot}^2+\kappa^2}}}{2 \kappa
  \sqrt{k_{\bot}^2+\kappa^2}} \nonumber \\ 
& \times \displaystyle\sum_{Q}D_{lmP,k_{\bot}Q}
r^{Q}D^{\dag}_{k_{\bot}Q,l'm'P'} \left(2 \delta_{Q,P'}-1\right),
\label{eq:N}
\end{align}
where $\vec{k}_{\bot}$ is the Fourier momentum parallel to the plate,
the $\mathcal{F}^{ee(j)}_{lmP,lmP}$ are the outside scattering
amplitudes of sphere $j$, $r^Q$ are the planar reflection coefficients
(Fresnel reflection coefficients in the case of an isotropic plate),
and $D_{lmP,k_{\bot m}}$ are conversion matrices:
\begin{align}
D_{lmE,k_{\bot}E} &= D_{lmM,k_{\bot}M} =
\sqrt{\frac{4\pi(2l+1)(l-m)!}{l(l+1)(l+m)!}}  \nonumber \\ &\times
\frac{|\vec{k}_{\bot}|}{\kappa} e^{-im\phi_{\vec{k}_{\bot}}}
P^{'m}_l\left(\sqrt{\vec{k}_{\bot}^2+\kappa^2}/\kappa\right)
\nonumber \\ D_{lmM,\vec{k}_{\bot}E} &= -D_{lmE,k_{\bot}M} =
-im\sqrt{\frac{4\pi(2l+1)(l-m)!}{l(l+1)(l+m)!}}  \nonumber \\ &\times
\frac{\kappa}{\vec{k}_{\bot}} e^{-im\phi_{\vec{k}_{\bot}}}
P_l^m\left(\sqrt{\vec{k}_{\bot}^2+\kappa^2}/\kappa\right),
\label{eq:Dmats}
\end{align}
given in terms of associated Legendre polynomials $P_l^m$ and their
derivatives with respect to their corresponding argument
$P^{'m}_l$. 

Upon a number of algebraic manipulations, similar expressions can be
obtained for the matrices $\mathcal{A}$ and $\mathcal{B}$, not found
in previous works, and in particular we find that:
\begin{align}
\label{eq:A}
  -\mathcal{A}_{lmP,l'm'P'}
  &=\mathcal{F}^{ee}_{R,lmP,lmP}\mathcal{U}^{23}_{lmP,l'm'P'}
  \nonumber \\ &+ \left(-1\right)^{m'-m}
  i^{m'-m}\mathcal{F}^{ee}_{R,lmP,lmP} \beta_{lmP, l'm' P'}
  \\ -\mathcal{B}_{lmP,l'm'P'} &=
  \mathcal{F}^{ee}_{L,lmP,lmP}\mathcal{U}^{32}_{lmP,l'm'P'} \nonumber
  \\ &+ i^{m'-m}\mathcal{F}^{ee}_{L,lmP,lmP} \beta_{lmP.l'm'P'},
\label{eq:B}
\end{align}
where
\begin{align}
\beta_{lmP,l'm'P'} &= \displaystyle\int_{0}^{\infty}
\frac{k_{\bot}dk_{\bot}}{(2\pi)} J_{m'-m}\left(S k_{\bot}\right)
\nonumber \\ & \times
\frac{e^{-(h_2+h_3)\sqrt{k_{\bot}^2+\kappa^2}}}{2\kappa
  \sqrt{k_{\bot}^2+\kappa^2}} \nonumber \\ & \times
\displaystyle\sum_{Q} D_{lmP,k_{\bot}Q} r^{Q}
D^{\dag}_{k_{\bot}Q,l'm'P'} \left(2 \delta_{Q,P'}-1\right),
\end{align}
and where the $J_m(Sk_{\bot})$ is a Bessel function of the first kind
evaluated at different values of $Sk_{\bot}$, where $S$ is given by the
projection of the sphere center--center separation onto the plate
axis:
\begin{equation}
  S=\sqrt{(d+R_1+R_2)^2-(h_1+R_1-h_2-R_2)^2}
\end{equation}

From a numerical perspective, all that remains in order to obtain the
Casimir energy in \eqref{caseq} is to evaluate the various matrix entries
and perform standard numerical operations, such as inversion and
multiplication, which we perform using standard free software
\cite{GSL}. For the small matrices that we consider, most of the time is
spent evaluating the various matrix elements, which can be numerically
expensive due to the integration of the oscillatory Bessel functions
in $\mathcal{A}$ and $\mathcal{B}$, although specialized methods for
oscillatory and Bessel integrals are available that may accelerate the
calculation~\cite{Evans:1999,Xiang:2007}.

\bibliographystyle{apsrev} 
\bibliography{photon}

\end{document}